\documentclass[twocolumn,english,prl,aps,english,superscriptaddress]{revtex4}
\usepackage[T1]{fontenc}
\usepackage[latin9]{inputenc}
\usepackage{color}
\usepackage{amsmath}
\usepackage{amssymb}
\usepackage{cancel}
\usepackage{graphicx}
\usepackage{esint}

\makeatletter

\@ifundefined{textcolor}{}
{%
 \definecolor{BLACK}{gray}{0}
 \definecolor{WHITE}{gray}{1}
 \definecolor{RED}{rgb}{1,0,0}
 \definecolor{GREEN}{rgb}{0,1,0}
 \definecolor{BLUE}{rgb}{0,0,1}
 \definecolor{CYAN}{cmyk}{1,0,0,0}
 \definecolor{MAGENTA}{cmyk}{0,1,0,0}
 \definecolor{YELLOW}{cmyk}{0,0,1,0}
}

\@ifundefined{textcolor}{}{%
 \definecolor{BLACK}{gray}{0}
 \definecolor{WHITE}{gray}{1}
 \definecolor{RED}{rgb}{1,0,0}
 \definecolor{GREEN}{rgb}{0,1,0}
 \definecolor{BLUE}{rgb}{0,0,1}
 \definecolor{CYAN}{cmyk}{1,0,0,0}
 \definecolor{MAGENTA}{cmyk}{0,1,0,0}
 \definecolor{YELLOW}{cmyk}{0,0,1,0}
}

\usepackage{dcolumn}
\usepackage{bm}
\usepackage{enumerate}

\usepackage{babel}

\usepackage{babel}

\usepackage{babel}

\usepackage{babel}

\makeatother

\usepackage{babel}
\begin{document}

\title{Phonon dynamics in correlated quantum systems driven away from equilibrium}

\author{Eli Y. Wilner}

\affiliation{School of Physics and Astronomy, The Sackler Faculty of Exact Sciences,
Tel Aviv University,Tel Aviv 69978,Israel}

\author{Haobin Wang}

\affiliation{Department of Chemistry, University of Colorado Denver, Denver, CO
80217-3364, USA}

\author{Michael Thoss}

\affiliation{Institute for Theoretical Physics and Interdisciplinary Center for
Molecular Materials, Friedrich-Alexander-Universit{ä}t Erlangen-N{ü}rnberg,
Staudtstr. 7/B2, 91058 Erlangen, Germany}

\author{Eran Rabani}

\affiliation{School of Chemistry, The Sackler Faculty of Exact Sciences, Tel Aviv
University,Tel Aviv 69978,Israel\\
 }
\begin{abstract}
A general form of a many-body Hamiltonian is considered, which includes
an interacting fermionic sub-system coupled to non-interacting extended
fermionic and bosonic systems. We show that the exact dynamics of
the extended bosonic system can be derived from the reduced density
matrix of the sub-system alone, despite the fact that the latter contains
information about the sub-system only. The advantage of the formalism
is immediately clear: While the reduced density matrix of the sub-system
is readily available, the formalism offers access to observables contained
in the full density matrix, which is often difficult to obtain. As
an example, we consider an extended Holstein model and study the nonequilibrium
dynamics of the, so called, ``reaction mode'' for different model
parameters. The effects of the phonon frequency, the strength of the
electron-phonon couplings, and the source-drain bias voltage on the
phonon dynamics across the bistability are discussed.
\end{abstract}
\maketitle
Strongly correlated systems show a remarkable range of interesting
phenomena, some of which can be explained with the current arsenal
of theoretical tools~\cite{strongly_corr_sys_review,tsvelik2001new}.
Their behavior when driven away from equilibrium (e.g., by a finite
bias voltage) is less well understood~\cite{stefanucci_nonequilibrium_2013}.
This is because theoretical tools that provide a reliable and accurate
description under equilibrium conditions are difficult to converge
for open quantum systems driven away from equilibrium~\cite{schmitteckert_nonequilibrium_2004,anders_real-time_2005}.
For example, Kondo physics in equilibrium has been understood within
renormalization group theory for several decades~\cite{wilson1975renormalization},
while the spectral properties under bias~\cite{Wingreen1994} were
only recently solved exactly by newly developed a numerical real-time
quantum Monte Carlo formalism~\cite{muhlbacher_real-time_2008,gull10_bold_monte_carlo},
confirming the voltage splitting of the Kondo peak~\cite{cohen2014green}.
The lack of a robust theoretical framework to nonequilibrium many-body
physics has been the driving force for developing theoretical tools
to understand both the dynamics and the approach to steady-state when
strong correlations are dominant.

One such powerful tool is based on the Nakjima--Zwanzig reduced density
matrix (RDM) formalism \cite{nakajima_quantum_1958,zwanzig_ensemble_1960,zwanzig_nonequilibrium_2001}
combined with a proper impurity solver to obtain the memory kernel~\cite{cohen_memory_2011}.
This approach has been applied recently to study charge and spin relaxation
near the Kondo crossover temperature of the Anderson impurity model~\cite{Cohen2013kondo}
and to study localization and bistability in the nonequilibrium extended
Holstein model~\cite{wilner_bistability_2013,wilner2014}. The Nakjima--Zwanzig
formalism, by construction, provides access to the dynamics of observables
within the reduced space only. Here, we expand the methodology and
show how to extract the dynamics of a certain class of observables
that were traced out. Our approach is particularly suitable for systems
with strong electron-phonon couplings, which give rise to highly interesting
phenomena~\cite{Blum2005,Sapmaz2006,Leturcq09,Lassange2009,Secker11}.
In light of this, we apply the formalism to analyze nonequilibrium
phonon dynamics in the extended Holstein model. While the nonequilibrium
phonon distribution in the steady state of this model has been analyzed
in great detail (see e.g. \cite{galperin_inelastic_2004,Mitra2004,Koch05,Galperin07,Peskin2011}
and references therein), so far there are only very few theoretical
studies of time-dependent phonon dynamics, which all involve significant
approximations~\cite{Metelmann2011,Dundas2012,Schiller2012,Albrecht2013}.
The methodology presented in this paper facilitates a numerically
converged treatment of this nonequilibrium many-body problem. 

To outline the reduced density matrix formalism, consider a general
Hamiltonian for a many-body quantum system comprising bosons and fermions
\begin{equation}
H=H_{s}+H_{f}+H_{b}+V_{sf}+V_{sb},\label{eq:general_ham}
\end{equation}
where $H_{s}=\sum_{ij}\varepsilon_{ij}d_{i}^{\dagger}d_{j}+\sum_{ijnm}V_{ijnm}d_{i}^{\dagger}d_{j}^{\dagger}d_{n}d_{m}$
and $H_{f}=\sum_{kq}\varepsilon_{kq}c_{k}^{\dagger}c_{q}$ are the
interacting and non-interacting parts of the Hamiltonian for the fermionic
degrees of freedom and $H_{b}=\sum_{\alpha}\hbar\omega_{\alpha}\left(b_{\alpha}^{\dagger}b_{\alpha}+\frac{1}{2}\right)$
describes the bosonic degrees of freedom. The coupling between the
sub-space ``$s$'' and ``$f$'' is described by $V_{sf}$ and
often is chosen in the form of a hopping between sites, $\sum_{ik}\left(t_{ik}d_{i}^{\dagger}c_{k}+h.c.\right)$,
but the formalism developed below is not limited to this choice. The
coupling between the interacting fermions and bosons is taken to the
lowest order in dimensionless boson coordinates, $x_{\alpha}=\frac{1}{\sqrt{2}}\left(b_{\alpha}^{\dagger}+b_{\alpha}\right)$:
\begin{equation}
V_{sb}=\sum_{ij,\alpha}M_{ij}^{\alpha}d_{i}^{\dagger}d_{j}x_{\alpha}.\label{eq:V_e-p}
\end{equation}
Here, $d_{i}^{\dagger}$/$d_{i}$ and $c_{k}^{\dagger}/c_{k}$ are
fermionic creation/annihilation operators at site $i$ and $k$, respectively,
and $b_{\alpha}^{\dagger}$/$b_{\alpha}$ are bosonic creation/annihilation
operators for mode $\alpha$. The above many-body Hamiltonian is a
general form covering different generic models, e.g., Fermi-Bose Hubbard
model~\cite{hubbard1963}, the spin-boson model~\cite{leggett1987dynamics},
and the Anderson-Holstein quantum impurity model~\cite{anderson1961,holstein_studies_1959}.
Thus, developing an approach to extract the time-dependent solution
of this generic model is of great importance.

Using the projection operator $\mathcal{P}=\rho_{f}\left(0\right)\otimes\rho_{b}\left(0\right)Tr_{f,b}$,
where $\rho_{f}\left(0\right)\otimes\rho_{b}\left(0\right)$ is the
initial density matrix in the ``$b$'' and ``$f$'' sub-spaces
and the trace $Tr_{f,b}$ is performed only for these degrees of freedom,
one can derive an exact equation of motion for the RDM of sub-space
``$s$'' (referred to as the ``system'')~\cite{wilner_bistability_2013}:

\begin{equation}
i\hbar\frac{\partial}{\partial t}\sigma\left(t\right)=\mathcal{L}_{s}\sigma\left(t\right)+\vartheta\left(t\right)-\frac{i}{\hbar}\int_{0}^{t}d\tau\kappa\left(\tau\right)\sigma\left(t-\tau\right),\label{eq:sigma(t)}
\end{equation}
where $\sigma\left(t\right)=Tr_{f,b}\,\rho\left(t\right)$ and $\rho\left(t\right)$
is the full time-dependent density matrix obeying the Liouville--Von-Neumann
equation $\dot{\rho}=\frac{i}{\hbar}\left[H,\rho\right]$. In the
above equation, $\mathcal{L}_{s}=[H_{s},\cdots]$ is the system's
Liouvillian, 
\begin{equation}
\vartheta\left(t\right)=Tr_{f,b}\left\{ \mathcal{L}_{v}e^{-\frac{i}{\hbar}\mathcal{Q}\mathcal{L}t}\mathcal{Q}\rho\left(0\right)\right\} \label{eq:theta(t)}
\end{equation}
is a super-operator matrix, that depends on the choice of initial
conditions and $\mathcal{L}_{v}=[V_{sf}+V_{sb},\cdots]$. By construction,
$\vartheta\left(t\right)$ vanishes for an uncorrelated initial state~\cite{wilner_bistability_2013},
\textit{i.e. }when $\rho(0)=\sigma(0)\otimes\rho_{f}\left(0\right)\otimes\rho_{b}\left(0\right)$,
where $\sigma(0)$ is the system initial density matrix. The memory
kernel super-operator, $\kappa\left(\tau\right),$ which describes
the non-Markovian dependency of the time propagation of the system,
is given by~\cite{wilner_bistability_2013} 
\begin{equation}
\kappa\left(t\right)=Tr_{f,b}\left\{ \mathcal{L}_{v}e^{-\frac{i}{\hbar}\mathcal{Q}\mathcal{L}t}\mathcal{Q}\mathcal{L}\left(\rho_{f}\left(0\right)\otimes\rho_{b}\left(0\right)\right)\right\} \label{eq:memory-kernel}
\end{equation}
with $\mathcal{Q}=1-\mathcal{P}$.

The calculation of the RDM in Eq.(\ref{eq:sigma(t)}) requires as
input the knowledge of the time-dependent memory kernel. The difficulty
in solving the many-body quantum Liouville--Von-Neumann equation for
$\rho\left(t\right)$ is now shifted to obtaining $\kappa\left(t\right)$.
However, simplifications can be made and rely on the fact that often
the memory kernel is short-lived (the time scale is typically governed
by a large energy scale), \emph{i.e.}, the system ``forgets'' its
history rapidly~\cite{cohen_generalized_2013}. Therefore, one can
develop suitable quantum impurity solvers to calculate the memory
until it decays and obtain the dynamics of the RDM at all times using
Eq.~\eqref{eq:sigma(t)}. 

The Nakjima--Zwanzig formalism, by construction, provides access to
the dynamics of observables within the reduced space only. Observables
that depend also on non-system degrees of freedom ($\in f,b$) can,
in principle, be calculated by introducing additional sets of Nakjima--Zwanzig
like equations. For example, for open quantum systems, the current
which depends both on $s$ and $f$ operators requires the introduction
of an additional memory term with a longer decay time~\cite{cohen_generalized_2013}.
The drawback of this extended Nakjima--Zwanzig formalism for non-system
operators is that each observable requires the introduction of an
additional memory term, which is often difficult (or perhaps impossible)
to calculate.

We propose an alternative formalism suitable for a certain class of
observables that does not require any additional calculation of memory
terms or the inclusion of the boson degrees of freedom in the system
part. More specifically, we show that the time evolution of the expectation
values of the positions and momenta of the bosonic variables can be
obtained from the RDM (or from the lesser two-time Green function,
$G^{<}\left(t,t\right)$) of the system alone, despite the fact that
$\sigma\left(t\right)$ (or $G^{<}\left(t,t\right)$) does not contain
any information about the bosonic bath that was traced out. The derivation
given below is rather simple but the result is powerful. It offers
a way to extract information which is not directly accessible. We
illustrate the approach for the extended nonequilibrium Holstein model
and discuss the correlations between the phonon and electron dynamics.

Consider the Heisenberg equation of motion for $b_{\alpha}\left(t\right)$
and $b_{\alpha}^{\dagger}\left(t\right)$ generated by the general
Hamiltonian of Eq.~\eqref{eq:general_ham}:

\begin{eqnarray}
\dot{b}_{\alpha}\left(t\right) & = & -i\omega_{\alpha}b_{\alpha}\left(t\right)-\frac{i}{\sqrt{2}\hbar}\sum_{ij}M_{ij}^{\alpha}d_{i}^{\dagger}\left(t\right)d_{j}\left(t\right)\nonumber \\
\dot{b}_{\alpha}^{\dagger}\left(t\right) & = & i\omega_{\alpha}b_{\alpha}^{\dagger}\left(t\right)+\frac{i}{\sqrt{2}\hbar}\sum_{ij}M_{ij}^{\alpha}d_{i}^{\dagger}\left(t\right)d_{j}\left(t\right),
\end{eqnarray}
where the dimensionless position and momentum of each boson mode is
given by $x_{\alpha}\left(t\right)=\frac{1}{\sqrt{2}}\left(b_{\alpha}^{\dagger}\left(t\right)+b_{\alpha}\left(t\right)\right)$
and $p_{\alpha}\left(t\right)=\frac{i}{\sqrt{2}}\left(b_{\alpha}^{\dagger}\left(t\right)-b_{\alpha}\left(t\right)\right)$,
respectively. Taking the expectation value over the initial density
matrix, $\rho(0)$, we find that:

\begin{eqnarray}
\left\langle \dot{x}_{\alpha}\left(t\right)\right\rangle  & = & \omega_{\alpha}\left\langle p_{\alpha}\left(t\right)\right\rangle \label{eq:x and p}\\
\left\langle \dot{p}_{\alpha}\left(t\right)\right\rangle  & = & -\omega_{\alpha}\left\langle x_{\alpha}\left(t\right)\right\rangle -\frac{1}{\hbar}\sum_{ij}M_{ij}^{\alpha}\left\langle d_{i}^{\dagger}\left(t\right)d_{j}\left(t\right)\right\rangle ,\nonumber 
\end{eqnarray}
where $\left\langle \cdots\right\rangle \equiv Tr[\rho(0)\cdots].$
The expectation value of the site populations and coherences can be
expressed in terms of the RDM (for the same sake by the elements of
the Green function of the system) by 
\begin{eqnarray}
\left\langle d_{i}^{\dagger}d_{i}\right\rangle  & = & \sum_{\begin{array}[t]{c}
n_{1},\ldots n_{N}\end{array}}\sigma_{n_{1},\ldots n_{N},n_{1},\ldots n_{N}}\delta_{n_{i},1}\label{eq:didj}\\
\left\langle d_{i}^{\dagger}d_{j}\right\rangle  & = & \sum_{\begin{array}[t]{c}
n_{1},\ldots n_{N}\\
n_{1}^{'},\ldots n_{N}^{'}
\end{array}}\sigma_{n_{1},\ldots n_{N},n_{1},\ldots n_{N}}\delta_{n_{i},1}\delta_{n_{j},0}\delta_{n_{i}^{'},0}\delta_{n_{j}^{'},1}\nonumber 
\end{eqnarray}
Eqs.~\eqref{eq:x and p} and \eqref{eq:didj} imply that if the RDM
of the system is known the average positions and momenta of the boson
modes can be obtained by solving for Eq.~\eqref{eq:x and p} with
the RDM given as an input. This is the main result of this letter.
We now illustrate this for the extended Holstein model.

In this model, $H_{s}=\varepsilon d^{\dagger}d$ includes a single
level, $H_{f}=\sum_{k\in L,R}\varepsilon_{k}c_{k}^{\dagger}c_{k}$,
and $H_{b}=\sum_{\alpha}\hbar\omega_{\alpha}\left(b_{\alpha}^{\dagger}b_{\alpha}+\frac{1}{2}\right)$.
The coupling between the system and the fermionic and bosonic baths
is given by $V_{sf}=\sum_{k}t_{k}d^{\dagger}c_{k}+h.c.$ and $V_{fb}=d^{\dagger}d\sum_{\alpha}M_{\alpha}\left(b_{\alpha}^{\dagger}+b_{\alpha}\right)$,
respectively. $t_{k}$ and $M_{\alpha}$ determine the strength of
the hybridization and electron-phonon couplings, respectively. The
former is modeled by a tight-binding spectral density with an overall
coupling determined by $\Gamma$ while the latter is modeled by an
Ohmic spectral density $J\left(\omega\right)=\frac{\pi\hbar}{2}\eta\omega e^{-\frac{\omega}{\omega_{c}}}$,
where the dimensionless Kondo parameter, $\eta=\frac{2\lambda}{\hbar\omega_{c}}$,
determines the overall electron-phonon couplings, $\omega_{c}$ is
the characteristic phonon bath frequency, and $\lambda=\sum_{\alpha}\frac{M_{\alpha}^{2}}{\hbar\omega_{\alpha}}=\frac{1}{\pi}\int\frac{d\omega}{\omega}J(\omega)$
is the reorganization energy (or polaron shift), which also determines
the shifting of the dot energy upon charging.

The reduced density matrix was recently solved for this model \cite{wilner_bistability_2013,wilner2014}
by employing two different approaches to calculate the memory kernel
and solving Eq.~\eqref{eq:sigma(t)} for $\sigma\left(t\right)$:
(i) a two-time nonequilibrium Green function (NEGF)~\cite{myohanen_kadanoff-baym_2010}
method within the self-consistent Born approximation (SCBA)~\cite{wilner2014}
suitable for weak electron-phonon couplings and (ii) the multilayer
multiconfiguration time-dependent Hartree (ML-MCTDH)~\cite{Thoss03,Wang2009},
which is numerically exact but more demanding. The results obtained
in a wide range of parameters revealed dynamics on multiple timescales.
In addition to the short and intermediate timescales associated with
the separate electronic and phononic degrees of freedom, the electron-phonon
coupling introduces longer timescales related to the adiabatic or
nonadiabatic tunneling between the two charge states ($\langle d^{\dagger}d\rangle=1$
and $\langle d^{\dagger}d\rangle=0$). The analysis shows, furthermore,
that the value of the dot occupation may depend on the initial preparation
of the phonon degrees of freedom, suggesting the existence of bistability~\cite{gogolin_multistable_2002,galperin_inelastic_2004,albrecht_bistability_2012,wilner_bistability_2013,wilner2014}.
Intriguingly, the phenomenon of bistability persists even on timescales
longer than the adiabatic/nonadiabatic tunneling time.

\begin{figure}[t]
\includegraphics[width=8cm]{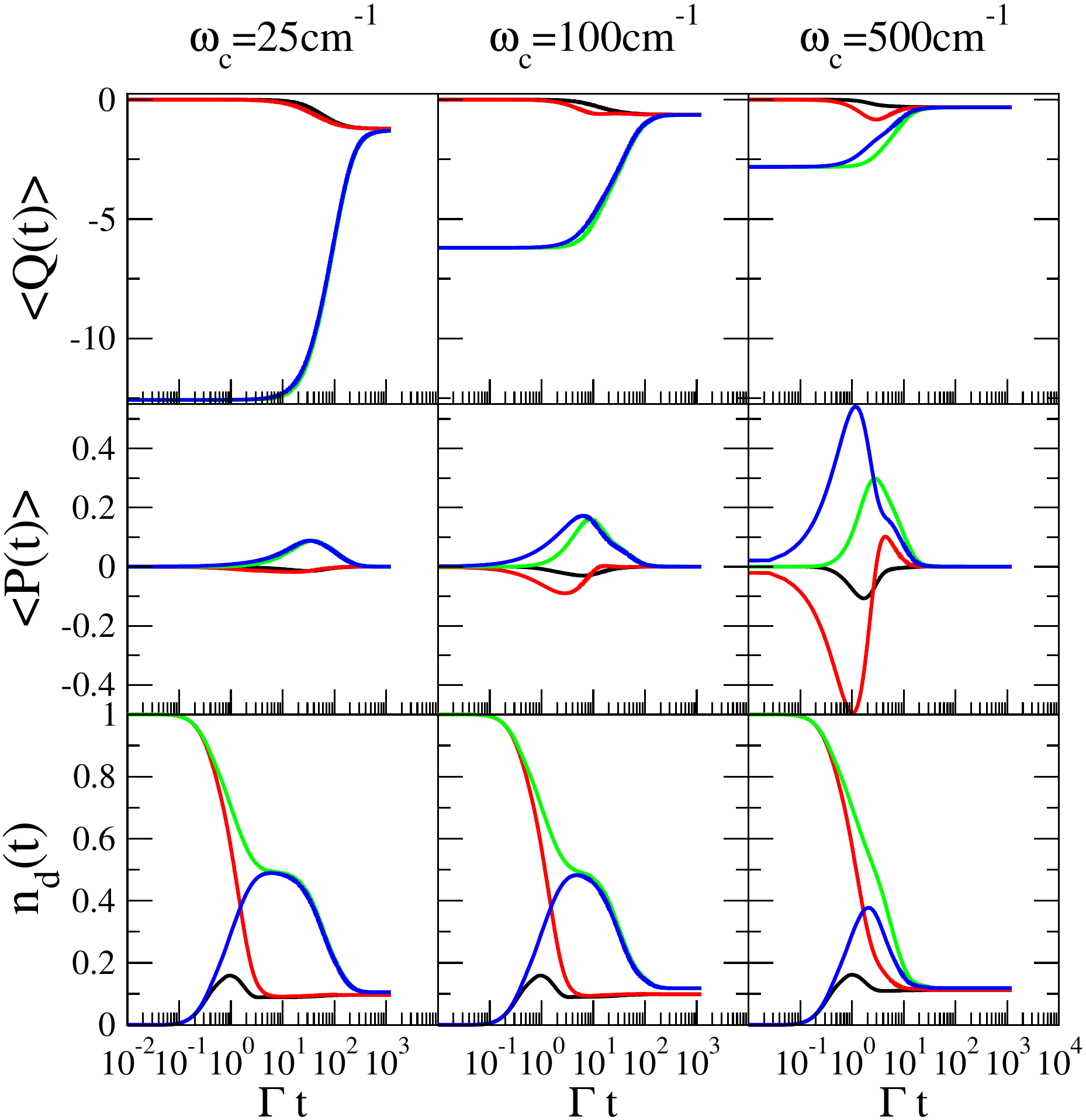} \protect\caption{\label{fig:epsilon_0.25_lamda_1000} $n_{d}\left(t\right)=\langle d^{\dagger}d\rangle(t),$
$\left\langle P\left(t\right)\right\rangle $, and $\left\langle Q\left(t\right)\right\rangle $
for different initial conditions: Black - unoccupied ($n_{d}\left(0\right)=0$)
with $\left\langle x_{\alpha}\left(0\right)\right\rangle =0$, \textcolor{red}{Red}
- occupied ($n_{d}\left(0\right)=1$) with $\left\langle x_{\alpha}\left(0\right)\right\rangle =0$,
\textcolor{blue}{Blue} - unoccupied with $\left\langle x_{\alpha}\left(0\right)\right\rangle =-\frac{2M_{\alpha}}{\hbar\omega_{\alpha}}$
and \textcolor{green}{Green} - occupied with $\left\langle x_{\alpha}\left(0\right)\right\rangle =-\frac{2M_{\alpha}}{\hbar\omega_{\alpha}}$.
The model parameters are $\lambda/\Gamma=0.77$, $\varepsilon_{d}/\Gamma=1.5625$,
$\Delta\mu/\Gamma=0.625$, and temperature $T=0$.}
\end{figure}

In Fig.~\ref{fig:epsilon_0.25_lamda_1000} we show the correlation
between the dynamics of the average dot occupation, the reaction mode
$\left\langle Q\left(t\right)\right\rangle =\sum_{\alpha}M_{\alpha}\left\langle x_{\alpha}\left(t\right)\right\rangle /\sqrt{\sum_{\alpha}M_{\alpha}^{2}}$,
and its corresponding momentum, $\left\langle P\left(t\right)\right\rangle =\left\langle \dot{Q}\left(t\right)\right\rangle /\Omega$,
where $\Omega=\frac{\int d\omega J\left(\omega\right)}{\int d\omega\frac{J\left(\omega\right)}{\omega}}=\omega_{c}$
is the reaction mode frequency. These results were obtained for weak
electron-phonon couplings by solving the memory kernel required to
obtain the RDM using the two-time NEGF with in the SCBA. Within this
limit, the NEGF--SCBA provides an accurate description of the RDM
in comparison to the numerically exact ML-MCTDH-SQR approach \cite{Thoss03,Wang2009}.
We consider $4$ different initial conditions for the system and boson
bath, namely all combinations of initial occupied/unoccupied ($n_{d}\left(0\right)=0/1$)
dot and shifted ($\left\langle x_{\alpha}\left(0\right)\right\rangle =-\frac{2M_{\alpha}}{\hbar\omega_{\alpha}}$)/unshifted
($\left\langle x_{\alpha}\left(0\right)\right\rangle =0$) phonon
modes. These shifted/unshifted values of $\left\langle x_{\alpha}\left(0\right)\right\rangle $
correspond to the location of the minimum of diabatic potential energy
of the occupied/unoccupied dot.

At long times, the dot population (lower panels of Fig.~\ref{fig:epsilon_0.25_lamda_1000})
decays to a unique value (closer to the empty state) regardless of
the initial preparation of the system and phonon bath, with a typical
decay time inversely proportional to $\Omega$ for the shifted bath
and to $\Gamma$ for the unshifted bath. The average position $\left\langle Q\left(t\right)\right\rangle $
and its corresponding momentum $\left\langle P\left(t\right)\right\rangle $
follow the population dynamics. At $t=0$, $\left\langle Q(0)\right\rangle $
assumes two different values corresponding to the left/right potential
minimum. Regardless of the initial conditions, the motion of the reaction
mode is overdamped (i.e. no oscillations are observed). This is known
to occur for the reaction mode of a bosonic bath with Ohmic spectral
density. At long times, the average position decays to values corresponding
to the more stable well, consistent with the behavior of the dot populations.
The typical time scale for approaching the steady-state value is given
by $\Omega$ (and not by $\Gamma$) regardless of the initial conditions
and varies only slightly for an unoccupied initial dot. 

\begin{figure}[t]
\includegraphics[width=8cm]{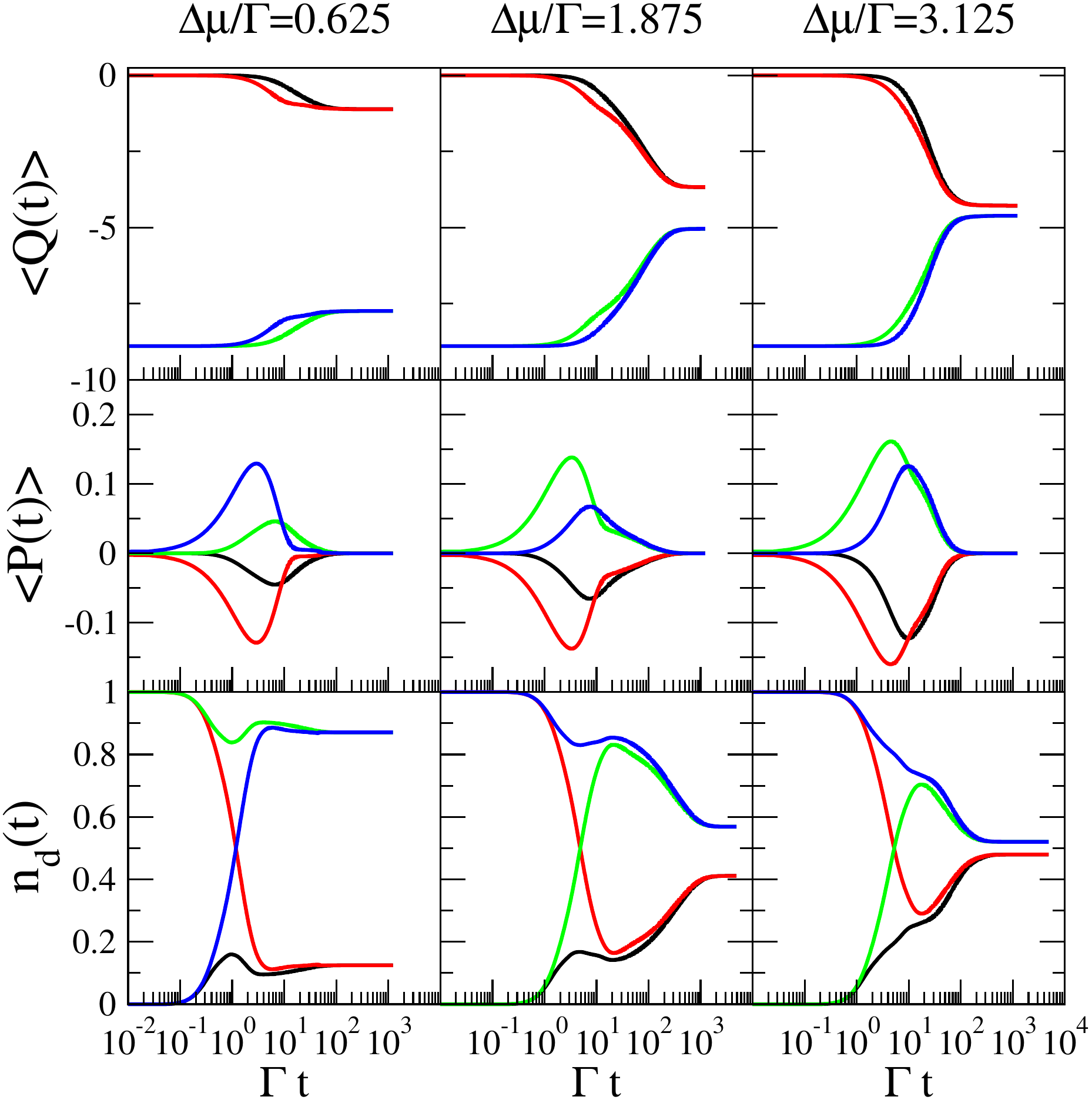} \protect\caption{\label{fig:epsilon_0.25_lamda_2000} Same as Fig.~\ref{fig:epsilon_0.25_lamda_1000}
for different values of the bias voltage. The model parameters are
$\lambda/\Gamma=\varepsilon_{d}/\Gamma=1.5625$ and $\omega_{c}=100\mbox{cm\ensuremath{^{-1}}}$.}
\end{figure}

The qualitative behavior of the dot population changes when the coupling
to the boson bath increases. In Fig.~\eqref{fig:epsilon_0.25_lamda_2000}
we show the results for a larger value of $\lambda=\varepsilon_{d}$
and different bias voltages $\Delta\mu$, still within the validity
of the NEGF-SCBA. In this case, both potential minima are degenerate
and the related spin-boson model (at equilibrium) shows a localization
transition at temperature $T=0$, which is broadened and finally disappears
as $T$ is increased. The appearance of two distinct values of the
dot population at long times at small bias voltages ($\Delta\mu$)
is consistent with the equilibrium results for the spin-boson model.
The bias voltage plays a similar role of temperature, and as its value
increases the bistability disappears.

Turning to discuss the transient behavior of $\left\langle Q\left(t\right)\right\rangle $
and $\left\langle P\left(t\right)\right\rangle $, we find that similar
to the previous case of weaker electron-phonon couplings, the average
position of the reaction coordinate follows closely the transient
behavior of the dot population. At long times $\left\langle Q\left(t\right)\right\rangle $
assumes two values roughly corresponding to the two minima of the
potential energy along the reaction mode, with vanishing differences
as $\Delta\mu$ increases. The corresponding average momenta always
decays to zero at long times, regardless of the initial conditions
of the dot and boson bath, suggesting that on the time scale observed
$\left\langle P\left(t\right)\right\rangle $ decays to its vanishing
steady-state value.

The relation between the behavior of $n_{d}(t)$ and $\left\langle Q\left(t\right)\right\rangle $
at steady state can be derived analytically. Since at steady-state
$\left\langle \dot{p}_{\alpha}\left(t\right)\right\rangle =0$, one
finds from Eq.~\eqref{eq:x and p} that each boson mode must satisfy
the relation $\left\langle x_{\alpha}\right\rangle =-\frac{M_{\alpha}}{\hbar\omega_{\alpha}}n_{d}$
and thus, the difference in $x_{\alpha}$ between the two different
initial distributions of the phonon modes is given by $M_{\alpha}\Delta x_{\alpha}=-\frac{M_{\alpha}^{2}}{\hbar\omega_{\alpha}}\Delta n_{d}$,
where $\Delta n_{d}$ is the corresponding difference between the
two dot populations at steady-state. Summing both sides over $\alpha$,
the reaction mode difference, $\Delta Q=\sum_{\alpha}M_{\alpha}\Delta x_{\alpha}/\sqrt{\sum_{\alpha}M_{\alpha}^{2}}$,
must satisfy the relation 
\begin{equation}
\Delta Q=-\frac{\lambda}{\sqrt{\sum_{\alpha}M_{\alpha}^{2}}}\Delta n_{d},\label{eq:delta_sigma_delta_q}
\end{equation}
where as before $\lambda=\sum_{\alpha}\frac{M_{\alpha}^{2}}{\hbar\omega_{\alpha}}$.
This is in agreement with the result obtained in Fig.~\ref{fig:epsilon_0.25_lamda_2000}.

\begin{figure}[t]
\includegraphics[width=8cm]{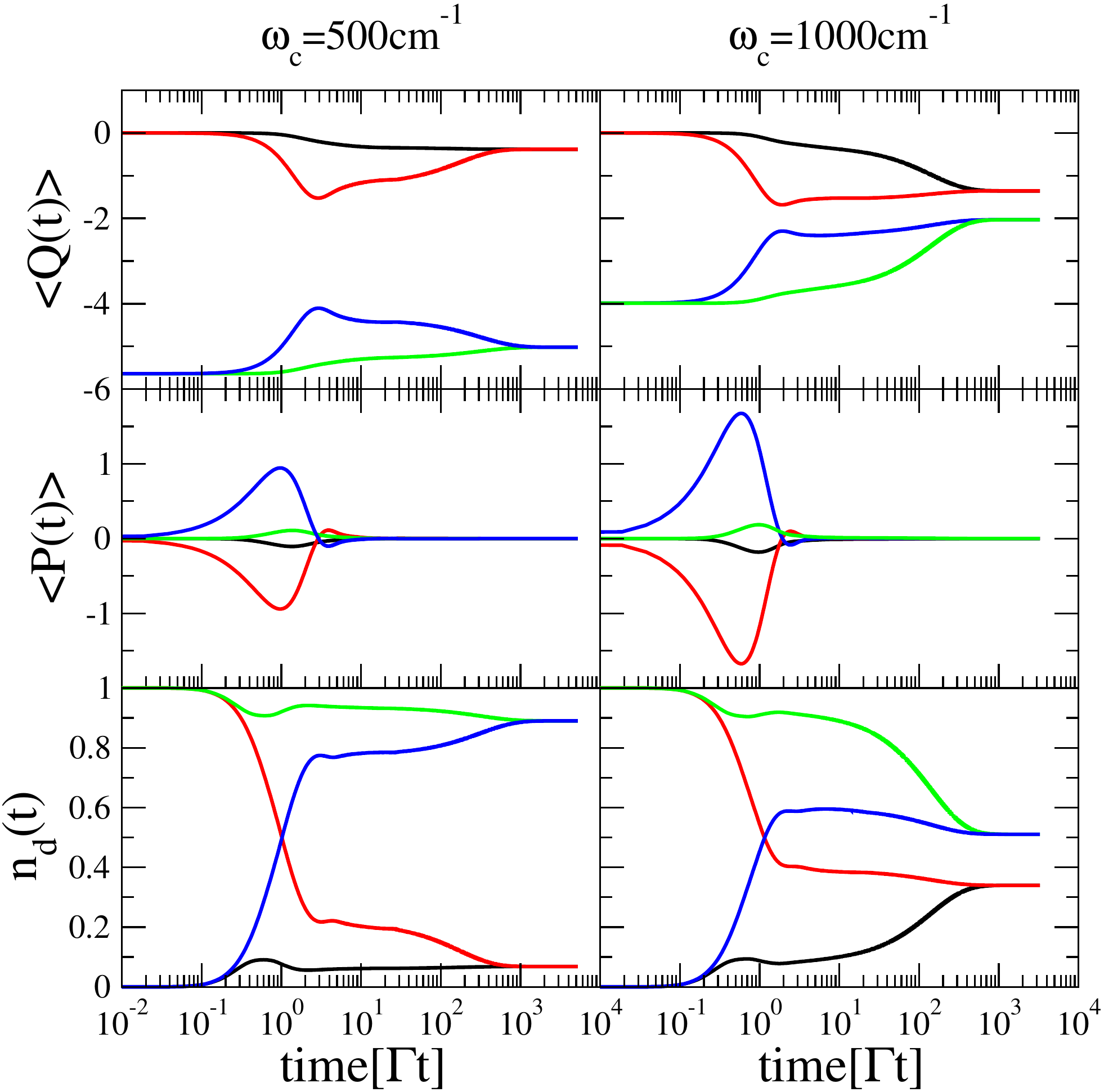}
\protect\caption{\label{fig:long-time} Same as Fig.~\ref{fig:epsilon_0.25_lamda_1000}
for $\lambda/\Gamma=3.1$ and $\varepsilon_{d}/\Gamma=3.125$. }
\end{figure}

So far, we have discussed the appearance of two bistable solutions
in the so called adiabatic limit, where $\omega_{c}\ll\Gamma$. In
Fig.~\ref{fig:long-time} we show results for larger values of $\omega_{c}$
in the regime where $\omega_{c}\approx\Gamma$. The value of the electron-phonon
coupling (reorganization energy) is somewhat larger than the perturbation
regime for which the NEGF-SCBA is accurate. Therefore, we obtain the
input required to generate the memory kernel and the RDM from the
numerically exact ML-MCTDH approach~\cite{Thoss03,Wang2009}. Similar
to the adiabatic limit with weaker electron-phonon couplings (shown
in Fig.~\ref{fig:epsilon_0.25_lamda_2000}), the long time behavior
of $n_{d}\left(t\right)$ depends on the initial conditions of the
phonon bath. However, unlike the adiabatic limit, here we find an
additional time scale at long times which is associated with the transition
from one diabatic potential surface to the other. Intriguingly, however,
the bistability prevails at times longer than the tunneling between
the two diabatic surfaces. As the phonon frequency increases, the
value of $\Delta n_{d}$ decreases and eventually disappears when
$\omega_{c}\gg\Gamma$.

Similar to the adiabatic limit, $Q\left(t\right)$ shows the same
behavior as $n_{d}\left(t\right)$, including the long time decay
associated with the aforementioned tunneling between the diabatic
surfaces, and the long time value of $\Delta Q$ is correlated with
that of $\Delta n_{d}$, in agreement with Eq.~\eqref{eq:delta_sigma_delta_q}.
Unlike the transient behavior of the reaction coordinate, its corresponding
momentum decays to its steady-state value on a much faster time scale
(typically on a time scale given by $\Omega^{-1}$) and does not show
the long time relaxation behavior associated with the phonon tunneling.
This implies that the tunneling process is not driven by inertia,
but is rather in the over-damped limit.

In summary, we have expanded our recently developed nonequilibrium
quantum dynamics methodology, which combines reduced density matrix
theory with an impurity solver to obtain the memory kernel, to describe
phonon dynamics in correlated open quantum systems. Although the phonon
degrees of freedom are formally not part of the reduced system, the
structure of the equations of motion allows the calculation of phonon
observables based solely on the density matrix and memory kernel of
the reduced system. The application to a Holstein-type model for phonon-coupled
electron transport in nanosystems reveals the intricate interplay
between electron and phonon dynamics in these systems, including the
phenomenon of bistability.

EYW is grateful to The Center for Nanoscience and Nanotechnology at
Tel Aviv University for a doctoral fellowship. HW acknowledges the
support from the National Science Foundation CHE-1361150. MT acknowledges
support from the German Research Foundation (DFG) and the German-Israeli
Foundation for Scientific Research and Development (GIF). This work
used resources of the National Energy Research Scientific Computing
Center, which is supported by the Office of Science of the U.S. Department
of Energy under Contract No. DE-AC02-05CH11231.


\begin{thebibliography}{41}
\expandafter\ifx\csname natexlab\endcsname\relax\def\natexlab#1{#1}\fi
\expandafter\ifx\csname bibnamefont\endcsname\relax
  \def\bibnamefont#1{#1}\fi
\expandafter\ifx\csname bibfnamefont\endcsname\relax
  \def\bibfnamefont#1{#1}\fi
\expandafter\ifx\csname citenamefont\endcsname\relax
  \def\citenamefont#1{#1}\fi
\expandafter\ifx\csname url\endcsname\relax
  \def\url#1{\texttt{#1}}\fi
\expandafter\ifx\csname urlprefix\endcsname\relax\def\urlprefix{URL }\fi
\providecommand{\bibinfo}[2]{#2}
\providecommand{\eprint}[2][]{\url{#2}}

\bibitem[{\citenamefont{Saxena and
  Littlewood}(2012)}]{strongly_corr_sys_review}
\bibinfo{author}{\bibfnamefont{S.~S.} \bibnamefont{Saxena}} \bibnamefont{and}
  \bibinfo{author}{\bibfnamefont{P.~B.} \bibnamefont{Littlewood}},
  \bibinfo{journal}{J. Phys.: Condens. Matter} \textbf{\bibinfo{volume}{24}},
  \bibinfo{pages}{290301} (\bibinfo{year}{2012}).

\bibitem[{\citenamefont{Tsvelik}(2001)}]{tsvelik2001new}
\bibinfo{author}{\bibfnamefont{A.}~\bibnamefont{Tsvelik}},
  \emph{\bibinfo{title}{New Theoretical Approaches to Strongly Correlated
  Systems}}, NATO science series: Mathematics, physics, and chemistry
  (\bibinfo{publisher}{Springer Netherlands}, \bibinfo{year}{2001}).

\bibitem[{\citenamefont{Stefanucci and
  Leeuwen}(2013)}]{stefanucci_nonequilibrium_2013}
\bibinfo{author}{\bibfnamefont{G.}~\bibnamefont{Stefanucci}} \bibnamefont{and}
  \bibinfo{author}{\bibfnamefont{R.~v.} \bibnamefont{Leeuwen}},
  \emph{\bibinfo{title}{Nonequilibrium Many-Body Theory of Quantum Systems: A
  Modern Introduction}} (\bibinfo{publisher}{Cambridge University Press},
  \bibinfo{year}{2013}), ISBN \bibinfo{isbn}{9780521766173}.

\bibitem[{\citenamefont{Schmitteckert}(2004)}]{schmitteckert_nonequilibrium_2004}
\bibinfo{author}{\bibfnamefont{P.}~\bibnamefont{Schmitteckert}},
  \bibinfo{journal}{Phys. Rev. B} \textbf{\bibinfo{volume}{70}},
  \bibinfo{pages}{121302} (\bibinfo{year}{2004}).

\bibitem[{\citenamefont{Anders and Schiller}(2005)}]{anders_real-time_2005}
\bibinfo{author}{\bibfnamefont{F.~B.} \bibnamefont{Anders}} \bibnamefont{and}
  \bibinfo{author}{\bibfnamefont{A.}~\bibnamefont{Schiller}},
  \bibinfo{journal}{Phys. Rev. Lett.} \textbf{\bibinfo{volume}{95}},
  \bibinfo{pages}{196801} (\bibinfo{year}{2005}).

\bibitem[{\citenamefont{Wilson}(1975)}]{wilson1975renormalization}
\bibinfo{author}{\bibfnamefont{K.~G.} \bibnamefont{Wilson}},
  \bibinfo{journal}{Rev. Mod. Phys} \textbf{\bibinfo{volume}{47}},
  \bibinfo{pages}{773} (\bibinfo{year}{1975}).

\bibitem[{\citenamefont{Wingreen and Meir}(1994)}]{Wingreen1994}
\bibinfo{author}{\bibfnamefont{N.~S.} \bibnamefont{Wingreen}} \bibnamefont{and}
  \bibinfo{author}{\bibfnamefont{Y.}~\bibnamefont{Meir}},
  \bibinfo{journal}{Phys. Rev. B} \textbf{\bibinfo{volume}{49}},
  \bibinfo{pages}{11040} (\bibinfo{year}{1994}).

\bibitem[{\citenamefont{M\"{u}hlbacher and
  Rabani}(2008)}]{muhlbacher_real-time_2008}
\bibinfo{author}{\bibfnamefont{L.}~\bibnamefont{M\"{u}hlbacher}}
  \bibnamefont{and} \bibinfo{author}{\bibfnamefont{E.}~\bibnamefont{Rabani}},
  \bibinfo{journal}{Phys. Rev. Lett.} \textbf{\bibinfo{volume}{100}},
  \bibinfo{pages}{176403} (\bibinfo{year}{2008}).

\bibitem[{\citenamefont{Gull et~al.}(2010)\citenamefont{Gull, Reichman, and
  Millis}}]{gull10_bold_monte_carlo}
\bibinfo{author}{\bibfnamefont{E.}~\bibnamefont{Gull}},
  \bibinfo{author}{\bibfnamefont{D.~R.} \bibnamefont{Reichman}},
  \bibnamefont{and} \bibinfo{author}{\bibfnamefont{A.~J.}
  \bibnamefont{Millis}}, \bibinfo{journal}{Phys. Rev. B}
  \textbf{\bibinfo{volume}{82}}, \bibinfo{pages}{075109}
  (\bibinfo{year}{2010}).

\bibitem[{\citenamefont{Cohen et~al.}(2014)\citenamefont{Cohen, Gull, Reichman,
  and Millis}}]{cohen2014green}
\bibinfo{author}{\bibfnamefont{G.}~\bibnamefont{Cohen}},
  \bibinfo{author}{\bibfnamefont{E.}~\bibnamefont{Gull}},
  \bibinfo{author}{\bibfnamefont{D.~R.} \bibnamefont{Reichman}},
  \bibnamefont{and} \bibinfo{author}{\bibfnamefont{A.~J.}
  \bibnamefont{Millis}}, \bibinfo{journal}{Phys. Rev. Lett}
  \textbf{\bibinfo{volume}{112}}, \bibinfo{pages}{146802}
  (\bibinfo{year}{2014}).

\bibitem[{\citenamefont{Nakajima}(1958)}]{nakajima_quantum_1958}
\bibinfo{author}{\bibfnamefont{S.}~\bibnamefont{Nakajima}},
  \bibinfo{journal}{Prog. Theor. Phys.} \textbf{\bibinfo{volume}{20}},
  \bibinfo{pages}{948} (\bibinfo{year}{1958}).

\bibitem[{\citenamefont{Zwanzig}(1960)}]{zwanzig_ensemble_1960}
\bibinfo{author}{\bibfnamefont{R.}~\bibnamefont{Zwanzig}}, \bibinfo{journal}{J.
  Chem. Phys.} \textbf{\bibinfo{volume}{33}}, \bibinfo{pages}{1338}
  (\bibinfo{year}{1960}).

\bibitem[{\citenamefont{Zwanzig}(2001)}]{zwanzig_nonequilibrium_2001}
\bibinfo{author}{\bibfnamefont{R.}~\bibnamefont{Zwanzig}},
  \emph{\bibinfo{title}{Nonequilibrium Statistical Mechanics}}
  (\bibinfo{publisher}{Oxford University Press}, \bibinfo{year}{2001}), ISBN
  \bibinfo{isbn}{9780195140187}.

\bibitem[{\citenamefont{Cohen and Rabani}(2011)}]{cohen_memory_2011}
\bibinfo{author}{\bibfnamefont{G.}~\bibnamefont{Cohen}} \bibnamefont{and}
  \bibinfo{author}{\bibfnamefont{E.}~\bibnamefont{Rabani}},
  \bibinfo{journal}{Phys. Rev. B} \textbf{\bibinfo{volume}{84}},
  \bibinfo{pages}{075150} (\bibinfo{year}{2011}).

\bibitem[{\citenamefont{Cohen et~al.}(2013{\natexlab{a}})\citenamefont{Cohen,
  Gull, Reichman, Millis, and Rabani}}]{Cohen2013kondo}
\bibinfo{author}{\bibfnamefont{G.}~\bibnamefont{Cohen}},
  \bibinfo{author}{\bibfnamefont{E.}~\bibnamefont{Gull}},
  \bibinfo{author}{\bibfnamefont{D.~R.} \bibnamefont{Reichman}},
  \bibinfo{author}{\bibfnamefont{A.~J.} \bibnamefont{Millis}},
  \bibnamefont{and} \bibinfo{author}{\bibfnamefont{E.}~\bibnamefont{Rabani}},
  \bibinfo{journal}{Phys. Rev. B} \textbf{\bibinfo{volume}{87}},
  \bibinfo{pages}{195108} (\bibinfo{year}{2013}{\natexlab{a}}).

\bibitem[{\citenamefont{Wilner et~al.}(2013)\citenamefont{Wilner, Wang, Cohen,
  Thoss, and Rabani}}]{wilner_bistability_2013}
\bibinfo{author}{\bibfnamefont{E.~Y.} \bibnamefont{Wilner}},
  \bibinfo{author}{\bibfnamefont{H.}~\bibnamefont{Wang}},
  \bibinfo{author}{\bibfnamefont{G.}~\bibnamefont{Cohen}},
  \bibinfo{author}{\bibfnamefont{M.}~\bibnamefont{Thoss}}, \bibnamefont{and}
  \bibinfo{author}{\bibfnamefont{E.}~\bibnamefont{Rabani}},
  \bibinfo{journal}{Phys. Rev. B} \textbf{\bibinfo{volume}{88}},
  \bibinfo{pages}{045137} (\bibinfo{year}{2013}).

\bibitem[{\citenamefont{Wilner et~al.}(2014)\citenamefont{Wilner, Wang, Thoss,
  and Rabani}}]{wilner2014}
\bibinfo{author}{\bibfnamefont{E.~Y.} \bibnamefont{Wilner}},
  \bibinfo{author}{\bibfnamefont{H.}~\bibnamefont{Wang}},
  \bibinfo{author}{\bibfnamefont{M.}~\bibnamefont{Thoss}}, \bibnamefont{and}
  \bibinfo{author}{\bibfnamefont{E.}~\bibnamefont{Rabani}},
  \bibinfo{journal}{Phys. Rev. B} \textbf{\bibinfo{volume}{89}},
  \bibinfo{pages}{205129} (\bibinfo{year}{2014}).

\bibitem[{\citenamefont{Blum et~al.}(2005)\citenamefont{Blum, Kushmerick, Long,
  Patterson, Yang, Henderson, Yao, Tour, Shashidhar, and R.Ratna}}]{Blum2005}
\bibinfo{author}{\bibfnamefont{A.~S.} \bibnamefont{Blum}},
  \bibinfo{author}{\bibfnamefont{J.~G.} \bibnamefont{Kushmerick}},
  \bibinfo{author}{\bibfnamefont{D.~P.} \bibnamefont{Long}},
  \bibinfo{author}{\bibfnamefont{C.~H.} \bibnamefont{Patterson}},
  \bibinfo{author}{\bibfnamefont{J.~C.} \bibnamefont{Yang}},
  \bibinfo{author}{\bibfnamefont{J.~C.} \bibnamefont{Henderson}},
  \bibinfo{author}{\bibfnamefont{Y.}~\bibnamefont{Yao}},
  \bibinfo{author}{\bibfnamefont{J.~M.} \bibnamefont{Tour}},
  \bibinfo{author}{\bibfnamefont{R.}~\bibnamefont{Shashidhar}},
  \bibnamefont{and} \bibinfo{author}{\bibfnamefont{B.}~\bibnamefont{R.Ratna}},
  \bibinfo{journal}{Nat. Mater.} \textbf{\bibinfo{volume}{4}},
  \bibinfo{pages}{167} (\bibinfo{year}{2005}).

\bibitem[{\citenamefont{Sapmaz et~al.}(2006)\citenamefont{Sapmaz,
  Jarillo-Herrero, Blanter, Dekker, and {van der Zant}}}]{Sapmaz2006}
\bibinfo{author}{\bibfnamefont{S.}~\bibnamefont{Sapmaz}},
  \bibinfo{author}{\bibfnamefont{P.}~\bibnamefont{Jarillo-Herrero}},
  \bibinfo{author}{\bibfnamefont{Y.~M.} \bibnamefont{Blanter}},
  \bibinfo{author}{\bibfnamefont{C.}~\bibnamefont{Dekker}}, \bibnamefont{and}
  \bibinfo{author}{\bibfnamefont{H.~S.~J.} \bibnamefont{{van der Zant}}},
  \bibinfo{journal}{Phys. Rev. Lett.} \textbf{\bibinfo{volume}{96}},
  \bibinfo{pages}{026801} (\bibinfo{year}{2006}).

\bibitem[{\citenamefont{Leturcq et~al.}(2009)\citenamefont{Leturcq, Stampfer,
  Inderbitzin, Durrer, Hierold, Mariani, Schultz, {von Oppen}, and
  Ensslin}}]{Leturcq09}
\bibinfo{author}{\bibfnamefont{R.}~\bibnamefont{Leturcq}},
  \bibinfo{author}{\bibfnamefont{C.}~\bibnamefont{Stampfer}},
  \bibinfo{author}{\bibfnamefont{K.}~\bibnamefont{Inderbitzin}},
  \bibinfo{author}{\bibfnamefont{L.}~\bibnamefont{Durrer}},
  \bibinfo{author}{\bibfnamefont{C.}~\bibnamefont{Hierold}},
  \bibinfo{author}{\bibfnamefont{E.}~\bibnamefont{Mariani}},
  \bibinfo{author}{\bibfnamefont{M.}~\bibnamefont{Schultz}},
  \bibinfo{author}{\bibfnamefont{F.}~\bibnamefont{{von Oppen}}},
  \bibnamefont{and} \bibinfo{author}{\bibfnamefont{K.}~\bibnamefont{Ensslin}},
  \bibinfo{journal}{Nat. Phys.} \textbf{\bibinfo{volume}{5}},
  \bibinfo{pages}{327} (\bibinfo{year}{2009}).

\bibitem[{\citenamefont{Lassange et~al.}(2009)\citenamefont{Lassange,
  Tarakanov, Kiranet, Garcia-Sanchez, and Bachthold}}]{Lassange2009}
\bibinfo{author}{\bibfnamefont{B.}~\bibnamefont{Lassange}},
  \bibinfo{author}{\bibfnamefont{Y.}~\bibnamefont{Tarakanov}},
  \bibinfo{author}{\bibfnamefont{J.}~\bibnamefont{Kiranet}},
  \bibinfo{author}{\bibfnamefont{D.}~\bibnamefont{Garcia-Sanchez}},
  \bibnamefont{and}
  \bibinfo{author}{\bibfnamefont{A.}~\bibnamefont{Bachthold}},
  \bibinfo{journal}{Science} \textbf{\bibinfo{volume}{325}},
  \bibinfo{pages}{1107} (\bibinfo{year}{2009}).

\bibitem[{\citenamefont{Secker et~al.}(2011)\citenamefont{Secker, Wagner,
  {H\"artle}, Thoss, and Weber}}]{Secker11}
\bibinfo{author}{\bibfnamefont{D.}~\bibnamefont{Secker}},
  \bibinfo{author}{\bibfnamefont{S.}~\bibnamefont{Wagner}},
  \bibinfo{author}{\bibfnamefont{S.~B.~R.} \bibnamefont{{H\"artle}}},
  \bibinfo{author}{\bibfnamefont{M.}~\bibnamefont{Thoss}}, \bibnamefont{and}
  \bibinfo{author}{\bibfnamefont{H.~B.} \bibnamefont{Weber}},
  \bibinfo{journal}{Phys. Rev. Lett.} \textbf{\bibinfo{volume}{106}},
  \bibinfo{pages}{136807} (\bibinfo{year}{2011}).

\bibitem[{\citenamefont{Galperin et~al.}(2004)\citenamefont{Galperin, Ratner,
  and Nitzan}}]{galperin_inelastic_2004}
\bibinfo{author}{\bibfnamefont{M.}~\bibnamefont{Galperin}},
  \bibinfo{author}{\bibfnamefont{M.~A.} \bibnamefont{Ratner}},
  \bibnamefont{and} \bibinfo{author}{\bibfnamefont{A.}~\bibnamefont{Nitzan}},
  \bibinfo{journal}{J. Chem. Phys.} \textbf{\bibinfo{volume}{121}},
  \bibinfo{pages}{11965} (\bibinfo{year}{2004}).

\bibitem[{\citenamefont{Mitra et~al.}(2004)\citenamefont{Mitra, Aleiner, and
  Millis}}]{Mitra2004}
\bibinfo{author}{\bibfnamefont{A.}~\bibnamefont{Mitra}},
  \bibinfo{author}{\bibfnamefont{I.}~\bibnamefont{Aleiner}}, \bibnamefont{and}
  \bibinfo{author}{\bibfnamefont{A.}~\bibnamefont{Millis}},
  \bibinfo{journal}{Phys. Rev. B} \textbf{\bibinfo{volume}{69}},
  \bibinfo{pages}{245302} (\bibinfo{year}{2004}).

\bibitem[{\citenamefont{Koch and von Oppen}(2005)}]{Koch05}
\bibinfo{author}{\bibfnamefont{J.}~\bibnamefont{Koch}} \bibnamefont{and}
  \bibinfo{author}{\bibfnamefont{F.}~\bibnamefont{von Oppen}},
  \bibinfo{journal}{Phys. Rev. Lett.} \textbf{\bibinfo{volume}{94}},
  \bibinfo{pages}{206804} (\bibinfo{year}{2005}).

\bibitem[{\citenamefont{Galperin et~al.}(2007)\citenamefont{Galperin, Ratner,
  and Nitzan}}]{Galperin07}
\bibinfo{author}{\bibfnamefont{M.}~\bibnamefont{Galperin}},
  \bibinfo{author}{\bibfnamefont{M.~A.} \bibnamefont{Ratner}},
  \bibnamefont{and} \bibinfo{author}{\bibfnamefont{A.}~\bibnamefont{Nitzan}},
  \bibinfo{journal}{J. Phys.: Condens. Matter} \textbf{\bibinfo{volume}{19}},
  \bibinfo{pages}{103201} (\bibinfo{year}{2007}).

\bibitem[{\citenamefont{Volkovich et~al.}(2011)\citenamefont{Volkovich,
  H{\"a}rtle, Thoss, and Peskin}}]{Peskin2011}
\bibinfo{author}{\bibfnamefont{R.}~\bibnamefont{Volkovich}},
  \bibinfo{author}{\bibfnamefont{R.}~\bibnamefont{H{\"a}rtle}},
  \bibinfo{author}{\bibfnamefont{M.}~\bibnamefont{Thoss}}, \bibnamefont{and}
  \bibinfo{author}{\bibfnamefont{U.}~\bibnamefont{Peskin}},
  \bibinfo{journal}{Phys. Chem. Chem. Phys.} \textbf{\bibinfo{volume}{13}},
  \bibinfo{pages}{14333} (\bibinfo{year}{2011}).

\bibitem[{\citenamefont{Metelmann and Brandes}(2011)}]{Metelmann2011}
\bibinfo{author}{\bibfnamefont{A.}~\bibnamefont{Metelmann}} \bibnamefont{and}
  \bibinfo{author}{\bibfnamefont{T.}~\bibnamefont{Brandes}},
  \bibinfo{journal}{Phys. Rev. B} \textbf{\bibinfo{volume}{84}},
  \bibinfo{pages}{155455} (\bibinfo{year}{2011}).

\bibitem[{\citenamefont{L{\"u} et~al.}(2012)\citenamefont{L{\"u}, Brandbyge,
  ard T.~N.~Todorov, and Dundas}}]{Dundas2012}
\bibinfo{author}{\bibfnamefont{J.-T.} \bibnamefont{L{\"u}}},
  \bibinfo{author}{\bibfnamefont{M.}~\bibnamefont{Brandbyge}},
  \bibinfo{author}{\bibfnamefont{P.~H.} \bibnamefont{ard T.~N.~Todorov}},
  \bibnamefont{and} \bibinfo{author}{\bibfnamefont{D.}~\bibnamefont{Dundas}},
  \bibinfo{journal}{Phys. Rev. B} \textbf{\bibinfo{volume}{85}},
  \bibinfo{pages}{245444} (\bibinfo{year}{2012}).

\bibitem[{\citenamefont{Vinkler et~al.}(2012)\citenamefont{Vinkler, Schiller,
  and Andrei}}]{Schiller2012}
\bibinfo{author}{\bibfnamefont{Y.}~\bibnamefont{Vinkler}},
  \bibinfo{author}{\bibfnamefont{A.}~\bibnamefont{Schiller}}, \bibnamefont{and}
  \bibinfo{author}{\bibfnamefont{N.}~\bibnamefont{Andrei}},
  \bibinfo{journal}{Phys. Rev. B} \textbf{\bibinfo{volume}{85}},
  \bibinfo{pages}{035411} (\bibinfo{year}{2012}).

\bibitem[{\citenamefont{Albrecht et~al.}(2013)\citenamefont{Albrecht,
  Martin-Rodero, Monreal, M\"{u}hlbacher, and {Levy Yeyati}}}]{Albrecht2013}
\bibinfo{author}{\bibfnamefont{K.~F.} \bibnamefont{Albrecht}},
  \bibinfo{author}{\bibfnamefont{A.}~\bibnamefont{Martin-Rodero}},
  \bibinfo{author}{\bibfnamefont{R.~C.} \bibnamefont{Monreal}},
  \bibinfo{author}{\bibfnamefont{L.}~\bibnamefont{M\"{u}hlbacher}},
  \bibnamefont{and} \bibinfo{author}{\bibfnamefont{A.}~\bibnamefont{{Levy
  Yeyati}}}, \bibinfo{journal}{Phys. Rev. B} \textbf{\bibinfo{volume}{87}},
  \bibinfo{pages}{085127} (\bibinfo{year}{2013}).

\bibitem[{\citenamefont{Hubbard}(1963)}]{hubbard1963}
\bibinfo{author}{\bibfnamefont{J.}~\bibnamefont{Hubbard}},
  \bibinfo{journal}{Proc. Roy. Soc. Lon. A} \textbf{\bibinfo{volume}{276}},
  \bibinfo{pages}{238} (\bibinfo{year}{1963}).

\bibitem[{\citenamefont{Leggett et~al.}(1987)\citenamefont{Leggett,
  Chakravarty, Dorsey, Fisher, Garg, and Zwerger}}]{leggett1987dynamics}
\bibinfo{author}{\bibfnamefont{A.~J.} \bibnamefont{Leggett}},
  \bibinfo{author}{\bibfnamefont{S.}~\bibnamefont{Chakravarty}},
  \bibinfo{author}{\bibfnamefont{A.}~\bibnamefont{Dorsey}},
  \bibinfo{author}{\bibfnamefont{M.~P.} \bibnamefont{Fisher}},
  \bibinfo{author}{\bibfnamefont{A.}~\bibnamefont{Garg}}, \bibnamefont{and}
  \bibinfo{author}{\bibfnamefont{W.}~\bibnamefont{Zwerger}},
  \bibinfo{journal}{Rev. Mod. Phys.} \textbf{\bibinfo{volume}{59}},
  \bibinfo{pages}{1} (\bibinfo{year}{1987}).

\bibitem[{\citenamefont{Anderson}(1961)}]{anderson1961}
\bibinfo{author}{\bibfnamefont{P.~W.} \bibnamefont{Anderson}},
  \bibinfo{journal}{Phys. Rev.} \textbf{\bibinfo{volume}{124}},
  \bibinfo{pages}{41} (\bibinfo{year}{1961}).

\bibitem[{\citenamefont{Holstein}(1959)}]{holstein_studies_1959}
\bibinfo{author}{\bibfnamefont{T.}~\bibnamefont{Holstein}},
  \bibinfo{journal}{Ann. Phys.} \textbf{\bibinfo{volume}{8}},
  \bibinfo{pages}{325} (\bibinfo{year}{1959}).

\bibitem[{\citenamefont{Cohen et~al.}(2013{\natexlab{b}})\citenamefont{Cohen,
  Wilner, and Rabani}}]{cohen_generalized_2013}
\bibinfo{author}{\bibfnamefont{G.}~\bibnamefont{Cohen}},
  \bibinfo{author}{\bibfnamefont{E.~Y.} \bibnamefont{Wilner}},
  \bibnamefont{and} \bibinfo{author}{\bibfnamefont{E.}~\bibnamefont{Rabani}},
  \bibinfo{journal}{New J. Phys.} \textbf{\bibinfo{volume}{15}},
  \bibinfo{pages}{073018} (\bibinfo{year}{2013}{\natexlab{b}}).

\bibitem[{\citenamefont{My{\"o}h{\"a}nen
  et~al.}(2010)\citenamefont{My{\"o}h{\"a}nen, Stan, Stefanucci, and
  Leeuwen}}]{myohanen_kadanoff-baym_2010}
\bibinfo{author}{\bibfnamefont{P.}~\bibnamefont{My{\"o}h{\"a}nen}},
  \bibinfo{author}{\bibfnamefont{A.}~\bibnamefont{Stan}},
  \bibinfo{author}{\bibfnamefont{G.}~\bibnamefont{Stefanucci}},
  \bibnamefont{and} \bibinfo{author}{\bibfnamefont{R.~v.}
  \bibnamefont{Leeuwen}}, \bibinfo{journal}{J. Phys. Conf. Ser.}
  \textbf{\bibinfo{volume}{220}}, \bibinfo{pages}{012017}
  (\bibinfo{year}{2010}).

\bibitem[{\citenamefont{Wang and Thoss}(2003)}]{Thoss03}
\bibinfo{author}{\bibfnamefont{H.}~\bibnamefont{Wang}} \bibnamefont{and}
  \bibinfo{author}{\bibfnamefont{M.}~\bibnamefont{Thoss}}, \bibinfo{journal}{J.
  Chem. Phys.} \textbf{\bibinfo{volume}{119}}, \bibinfo{pages}{1289}
  (\bibinfo{year}{2003}).

\bibitem[{\citenamefont{Wang and Thoss}(2009)}]{Wang2009}
\bibinfo{author}{\bibfnamefont{H.}~\bibnamefont{Wang}} \bibnamefont{and}
  \bibinfo{author}{\bibfnamefont{M.}~\bibnamefont{Thoss}}, \bibinfo{journal}{J.
  Chem. Phys.} \textbf{\bibinfo{volume}{131}}, \bibinfo{pages}{024114}
  (\bibinfo{year}{2009}).

\bibitem[{\citenamefont{Gogolin and Komnik}(2002)}]{gogolin_multistable_2002}
\bibinfo{author}{\bibfnamefont{A.~O.} \bibnamefont{Gogolin}} \bibnamefont{and}
  \bibinfo{author}{\bibfnamefont{A.}~\bibnamefont{Komnik}},
  \bibinfo{journal}{{arXiv:cond-mat/0207513}}  (\bibinfo{year}{2002}).

\bibitem[{\citenamefont{Albrecht et~al.}(2012)\citenamefont{Albrecht, Wang,
  M{\"u}hlbacher, Thoss, and Komnik}}]{albrecht_bistability_2012}
\bibinfo{author}{\bibfnamefont{K.~F.} \bibnamefont{Albrecht}},
  \bibinfo{author}{\bibfnamefont{H.}~\bibnamefont{Wang}},
  \bibinfo{author}{\bibfnamefont{L.}~\bibnamefont{M{\"u}hlbacher}},
  \bibinfo{author}{\bibfnamefont{M.}~\bibnamefont{Thoss}}, \bibnamefont{and}
  \bibinfo{author}{\bibfnamefont{A.}~\bibnamefont{Komnik}},
  \bibinfo{journal}{Phys. Rev. B} \textbf{\bibinfo{volume}{86}},
  \bibinfo{pages}{081412} (\bibinfo{year}{2012}).

\end{thebibliography}
\end{document}